\documentclass[usenatbib]{mn2e}

\usepackage{psfig,url}

\footnotesize
\newdimen\digitwidth    
\setbox0=\hbox{\rm0}
\digitwidth=\wd0
\catcode`!=\active
\def!{\kern\digitwidth}
\normalsize

\title[Pulsars in the Magellanic Clouds]
{New limits on the population of normal and millisecond pulsars 
in the Large and Small Magellanic Clouds}
\author[J.P.~Ridley \& D.R.~Lorimer]
{J.P.~Ridley$^{1}$ and D.R.~Lorimer$^{1,2}$
\\
$^1$Department of Physics, West Virginia University, 
PO~Box~6315, Morgantown, WV~26506, USA\\
$^2$National Radio Astronomy Observatory, PO Box 2, 
Green Bank, WV~24944, USA\\
}
\let\leqslant=\leq 
\date{Accepted 2010 May 22. Received 2010 May 21; in original form 2010 March 3}
\begin{document}
\maketitle
\newcommand{\setthebls}{
}
\setthebls

\begin{abstract} 
We model the potentially observable populations of normal and
millisecond radio pulsars in the Large and Small Magellanic Clouds
(LMC and SMC) where the known population currently stands at 19 normal
radio pulsars.  Taking into account the detection thresholds of
previous surveys, and assuming optimal period and luminosity
distributions based on studies of Galactic pulsars, we estimate there
are $(1.79 \pm 0.20) \times 10^4$ and $(1.09 \pm 0.16) \times 10^4$
normal pulsars in the LMC and SMC respectively.  When we attempt to
correct for beaming effects, and the fraction of high-velocity pulsars
which escape the clouds, we estimate birth rates in both the LMC and
SMC to be comparable and in the range 0.5--1 pulsar per century.
Although higher than estimates for the rate of core-collapse
supernovae in the clouds, these pulsar birth rates are consistent with
historical supernova observations in the past 300 yr.  
A substantial population of active radio pulsars (of order a few
hundred thousand) have escaped the LMC and SMC and populate the 
local intergalactic medium. For the millisecond pulsar (MSP) 
population, the lack of any detections from current surveys leads 
to respective upper limits (at the 95\% confidence level) of 15,000 
for the LMC and 23,000 for the SMC.  Several MSPs could be detected 
by a currently ongoing survey of the SMC with improved time and 
frequency resolution using the Parkes multibeam system. Giant-pulse 
emitting neutron stars could also be seen by this survey.
\end{abstract}

\begin{keywords}
pulsars:general
\end{keywords}

\section{INTRODUCTION}\label{sec:intro}

Our closest neighbouring galaxies, the Large and Small Magellanic
Clouds (henceforth LMC and SMC), contain 19 known radio
pulsars \citep{mfl+06}. Since both the LMC and SMC represent a
different star formation environment compared to our Galaxy and its
globular cluster system, it is interesting to use these results to
constrain the pulsar content in the clouds. So far, no millisecond 
pulsars (MSPs) have been found, though this is in part a result of
the relatively low sampling rates and frequency resolutions of
previous surveys (see Table~\ref{tb:param}). The aim of this paper 
is to use Monte Carlo simulations to reproduce the results from 
previous LMC and SMC surveys, determine the most likely size of the 
normal pulsar population, place upper limits on the MSP population,
and make predictions for future surveys of the clouds.

In Section~\ref{sec:surveys}, we compare some of the past surveys of
the Magellanic Clouds.  Section~\ref{sec:normal} contains a detailed 
description of our simulations and how we are able to reproduce results 
from the past surveys.  From this model, we are able to infer the 
population of normal pulsars in the LMC and SMC.  New limits from our 
MSP simulations are presented in Section~\ref{sec:msp}.   We discuss 
our results and present predictions for future Magellanic Cloud surveys 
in Section~\ref{sec:discussion}, and finally, in 
Section~\ref{sec:conclusions}, we draw our conclusions.

\section{PREVIOUS SURVEYS OF THE MAGELLANIC CLOUDS}\label{sec:surveys}

Multiple radio surveys have been completed of both the Large and Small 
Magellanic clouds. We discuss the various features and results of each 
survey in this section. The survey data acquisition parameters
are summarized in Table~\ref{tb:param}.

\begin{table*}
\caption{Survey parameters for the previous surveys considered in this paper, 
as well as the parameters for a currently ongoing survey using the BPSR 
pulsar backend at the Parkes Radio Telescope.}
\label{tb:param}
\begin{tabular}{lcccc}
\hline
Parameter & McConnell et al. (1991) & Crawford et al. (2001) & Manchester 
et al. (2006) & BPSR Survey (2009) \\
\hline
$t_{\rm obs}$ (s)       	& 5000  & 8400  & 8400  & 8400\\
$t_{\rm samp}$ (ms) 		& 5.000 & 0.250 & 1.000 & 0.064\\
$T_{\rm sys}$ (K)      		& 60    & 21    & 21    & 20\\
$f$ (MHz)       		& 610   & 1374  & 1374  & 1374\\
$\Delta f$ (MHz)		& 120   & 288   & 288   & 288\\
$\Delta f_{\rm chan}$ (MHz)	& 2.5   & 3.0   & 3.0 & 0.1\\
\hline
\end{tabular}
\end{table*}

The original pulsar survey of the clouds was by \citet{mmh+91} which
detected three pulsars in the LMC and one in the SMC.  With the
advent of the Parkes Multibeam Pulsar Survey \citep{mlc+01}, searches
of the clouds with greatly increased sensitivity were
performed. \citet{ckm+01} surveyed the SMC, finding two more associated
pulsars, one in each cloud.  \citet{mfl+06} extended this
effort, surveying both the SMC and LMC. For the SMC, this survey
discovered 3 new pulsars, while the LMC survey discovered 9 more normal 
pulsars.  Currently, there are a total of 19 known spin-powered radio
pulsars in the LMC and SMC.

\begin{table*} 
\caption{List of all spin-powered radio pulsars in the LMC and SMC.  
We note that the Crawford et al. survey did not observe the entire 
LMC, so we do not consider their LMC results in our simulations.
Also, PSR B0540-69 is the only currently known LMC pulsar that 
was not detected in any of these surveys.}
\label{tb:psrlist}
\begin{tabular}{lcrccc}
\hline
PSR Name   & Location & DM (pc cm$^{-3}$) & McConnell et al. (1991) & Crawford et al. (2001) & 
Manchester et al. (2006) \\
\hline
J0045$-$7319 & SMC & 105.4 & X & X & X\\
J0455$-$6951 & LMC & 94.9  & X &   & X\\
J0502$-$6617 & LMC & 68.9  & X &   & X\\
J0529$-$6652 & LMC & 103.2 & X &   & X\\
J0113$-$7220 & SMC & 125.5 &   & X & X\\
J0535$-$6935 & LMC & 93.7  &   & X &  \\
B0540$-$69   & LMC & 146.5 &   &   &  \\
J0045$-$7042 & SMC & 70.0  &   &   & X\\
J0111$-$7131 & SMC & 76.0  &   &   & X\\
J0131$-$7310 & SMC & 205.2 &   &   & X\\
J0449$-$7031 & LMC & 65.8  &   &   & X\\
J0451$-$67   & LMC & 45.0  &   &   & X\\
J0456$-$7031 & LMC & 100.3 &   &   & X\\
J0519$-$6932 & LMC & 119.4 &   &   & X\\
J0522$-$6847 & LMC & 126.5 &   &   & X\\
J0532$-$6639 & LMC & 69.3  &   &   & X\\
J0534$-$6703 & LMC & 94.7  &   &   & X\\
J0543$-$6851 & LMC & 131.0 &   &   & X\\
J0555$-$7056 & LMC & 73.4  &   &   & X\\
\hline
TOTALS	& SMC & & 1 & 2  & 5\\ 
	& LMC & & 3 & NA & 12\\	
\hline
\end{tabular}
\end{table*}

\section{MODELING THE NORMAL PULSAR POPULATION}\label{sec:normal}

Our goal is to create a working model of normal pulsars distributed
throughout the Magellanic Clouds that can accurately reproduce the
results found in each of the three surveys carried out so far.  To 
do this, we perform Monte Carlo simulations of the pulsar population
using an adapted version of the freely available {\tt PSRPOP} software
package\footnote{http://psrpop.sourceforge.net} \citep{lfl+06}.

The simulation begins by seeding each pulsar in the 
Magellanic Clouds.  To do this for the SMC, we consider 
a sphere with a radius of 2.0~kpc centred at a Right Ascension
of 0h~51m and a Declination of $-73^{\circ}~07'$.  We 
populate this sphere with pulsars using an exponential 
distribution (in the x, y, and z directions) having a 
scale length of 1.25~kpc \citep{glk+09}.
Similarly, the LMC simulations create a sphere with a 
radius of 4.5~kpc centred at 05h~17m and 
$-69^{\circ}~02'$ \citep{ksd+98}, and use an exponential 
distribution with a scale length of 1.50~kpc \citep{vdm01}.

The dispersion measure (DM) is randomly selected in the range 
70--210~cm$^{-3}$~pc.  This is based on the DMs of other pulsars 
found in the Magellanic Clouds listed in Table~\ref{tb:psrlist}.  
For the distance to each pulsar, we use a value of 51~kpc for the 
LMC \citep{k09} and 60~kpc for the SMC \citep{hhh03}. 

The period of each pulsar is then randomly generated from a log-normal
distribution derived from a study of Galactic pulsars \citep{lfl+06}.  
For a period in ms, the mean of this distribution is 2.7 and the
standard deviation is 0.34. Similarly, the radio luminosity 
is drawn from the log-normal function found by \citet{fk06}.

To determine the number of pulsars to generate, we use one set of
survey parameters and attempt to replicate their results.  For our
trials, we choose the \citet{mfl+06} survey because of its larger sample
of detected pulsars.  We create a sample distribution for both the LMC
and the SMC.  We run {\tt PSRPOP} and generate enough SMC pulsars so
that the Manchester survey would detect 5 of them, and enough LMC
pulsars so that it would detect 12.  We then use these two
distributions to analyse the McConnell and Crawford surveys.

Table~\ref{tb:expected} shows how the results of all three surveys
compared with what we expect to detect.  As can be seen, these results
provide a self-consistent model of the Magellanic Cloud pulsars.  This
allows us to now model any future survey using this distribution of
pulsars and predict what results should be obtained.

\begin{table}
\caption{
Survey results for the three surveys considered in our paper.  We
compare what was actually detected to what our sample pulsar
distributions predict to be detected.  The first number is the actual
number of pulsars detected, while the second number is the expected
number of detections.
}
\label{tb:expected}
\begin{center}
\begin{tabular}{lcc}
\hline
SURVEY & SMC & LMC \\
\hline
McConnell & 1/1 & 2/3 \\
Crawford & 2/2 & NA \\
Manchester & 5/5 & 12/12\\
\hline
\end{tabular}
\end{center}
\end{table}

\begin{figure}
\caption{
Distributions of the number of pulsars generated to model the
LMC normal pulsars (solid), alongside the SMC pulsars (dashed).
}
\label{fig:normal}
\psfig{file=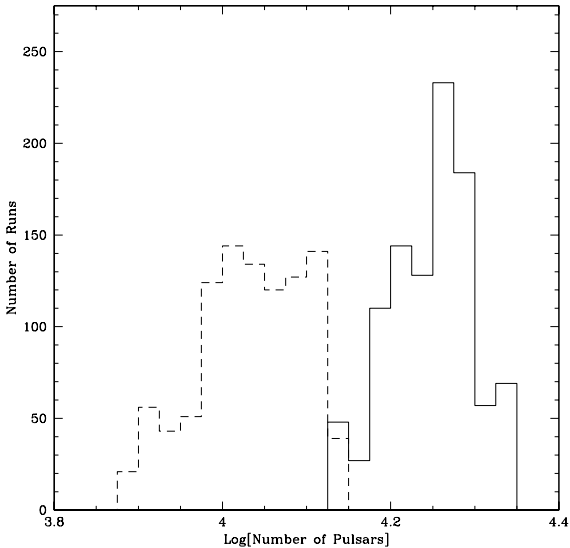,width=8cm,angle=0}
\end{figure}

To determine the number of normal pulsars in the LMC and SMC, we
generate enough pulsars to detect the corresponding number of pulsars
found in each cloud by the Manchester et al. survey.  Repeating this
process 1000 times allows us to get a mean value for each cloud.  The
LMC has a mean value of $(1.79 \pm 0.20) \times 10^4$ pulsars, while
the SMC has a mean of $(1.09 \pm 0.16) \times 10^4$.
Fig.~\ref{fig:normal} shows the distributions of normal pulsars
produced by these simulations.

\section{MODELING THE MILLISECOND PULSAR POPULATION}\label{sec:msp}

Since no MSPs have been detected in the Magellanic
Clouds, we can only find an upper limit on the number of MSPs
that could lie in the clouds.  As before, we use the Manchester et al.
survey as a reference, and we follow a procedure similar to that
described in the previous section.  We keep all pulsar parameters the
same, with the exception of the period distribution.  Here we use a
distribution of MSP periods used in a recent study 
\nocite{slk+09} (Smits et al. 2009).  For each of the LMC and 
SMC, we again run {\tt PSRPOP} until the Manchester et al. survey 
detects one of the MSPs.  The number of pulsars generated to get 
that one detection will be the upper limit of the number of MSPs 
in the cloud.

\begin{figure}
\caption{
Similar to normal pulsar histograms, we show the distribution of pulsars 
generated for the MSP simulations of the LMC (solid) and SMC (dashed).
} 
\label{fig:msp}
\psfig{file=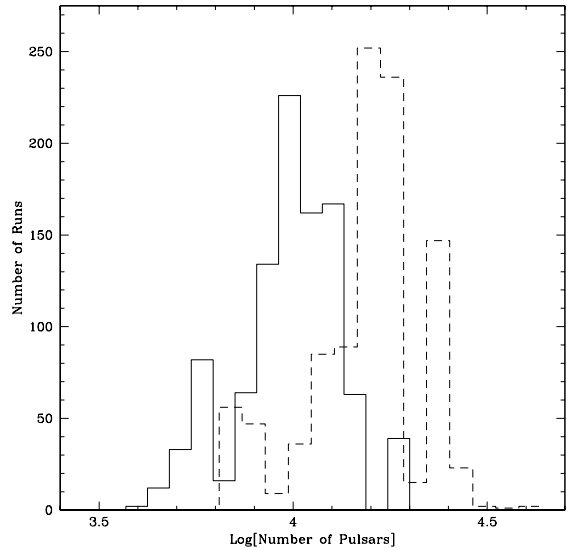,width=8cm,angle=0}
\end{figure}

To limit statistical noise and fluctuations, we run 1000
simulations, create a histogram of the number of pulsars
generated, and determine the mean.  To further limit these variances,
we run our simulation to detect 10 MSPs, 
and then divide our final number of pulsars generated by 10.  
Fig.~\ref{fig:msp} shows our simulation results.

To determine upper limits from these distributions, we consider a
point below which 95\% of our simulations are contained.  For the LMC
MSP population, we find, with 95\% confidence, an upper limit of
15,000 pulsars.  The equivalent simulations of the SMC yield somewhat
higher numbers, presumably due to the fact that they are located at a
farther distance from the Earth.  The corresponding 95\% confidence
upper limit is 23,000.

\section{DISCUSSION}\label{sec:discussion}

We have produced a self-consistent model of the normal pulsar
population in the LMC and SMC. We now make some straightforward
inferences from our results and comment on the yields of future
Magellanic Cloud surveys.

\subsection{Birthrates of normal pulsars}

To estimate the birthrate of normal pulsars in the 
Magellanic Clouds, we use the results \cite{fk06} to obtain an average 
lifetime for Galactic pulsars.  When their estimated number of pulsars of 
$(1.2 \pm 0.3) \times 10^6$ is divided by their Galactic pulsar 
birthrate of $2.8 \pm 0.1$ pulsars per century, we get an average radio
lifetime of $(4.3 \pm 0.1) \times 10^7$ years.  
To calculate the birthrates in the LMC and SMC, we divide the 
respective number of pulsars we find by this average lifetime, and then 
multiply by a factor to correct for those pulsars whose beams we do not
see due to beaming effects.  This factor is approximately 10 
when averaged over the normal pulsar population \citep{tm98}.
These calculations lead to birthrates of $0.42 \pm 0.03$ 
and $0.25 \pm 0.03$ pulsars per century for the LMC and SMC respectively.

These birth rates, however, are likely to be underestimates
of the true value due to the fact that both the LMC and SMC have 
relatively small escape velocities.  Using values for the mass and 
radius of the SMC from \citet{bs09}, we calculate an escape velocity
$v_{\rm esc} = \sqrt{2 G M/R}$, of roughly 120 km~s$^{-1}$.  
Similarly, we find an escape velocity for the LMC of 157 km~s$^{-1}$.  
Given the high birth velocities known for Galactic pulsars 
\citep[e.g.]{hllk05,fk06}, potentially a large number of pulsars 
that are born in the clouds will eventually lie outside of the 
surveyed area within the age in which they are expected to be radio loud. 

To quantify the impact of this effect on the birthrates,
we ran a simple simulation in which the velocity of each pulsar 
was chosen from a Maxwellian distribution with a 1-D dispersion of 
265~km~s$^{-1}$ \citep{hllk05}.  After randomly selecting the age of 
the pulsar to be between 0 and $4.3 \times 10^7$ years, we calculate 
the distance the pulsar has traveled by multiplying the age and the 
velocity.  Assuming that the pulsar starts at the centre of either 
cloud, we compute the final position to determine if the pulsar has 
remained within the surveyed area of the cloud (radii of $5.6^{\circ}$ 
and $2.2^{\circ}$ for the LMC and SMC) for the duration of 
its lifetime. We find that nearly half of all pulsars born in the 
LMC (45.9\%) escape the cloud, while 67.2\% of all pulsars born in 
the SMC escape. The population of ejected radio pulsars that are
active from both Magellanic Clouds is therefore substantial, of order
$3.5 \times 10^{5}$ objects.

The above calculations lead us to modified birthrates of 
0.8 pulsars per century for both the LMC and SMC. 
Given the uncertainties involved in making these corrections, the 
best conclusion we can reach from the data at hand is that radio 
pulsar birth rates are comparable and likely to be somewhere in 
the range of 0.5--1 pulsar per century in both the LMC and SMC.

Tammann, L\"offler \& Schr\"oder (1994) \nocite{tls94} carried out 
an extensive study of the Galactic and extragalactic supernova rate.
They find a combined rate of type Ib and II supernovae
of 0.45 and 0.11 per century in the LMC and SMC respectively.
While the rate for the LMC is consistent with our estimate above,
the pulsar birthrate in the SMC appears to be significantly higher
than these supernova rates. In spite of this discrepancy, a simple exercise using
Poissonian statistics shows that both pulsar and supernovae
populations are consistent with the observation of one supernova in the
LMC and zero supernovae in the SMC during the past 300 years of observations.

\subsection{Predictions for future surveys}\label{sec:bpsr}

Previous surveys were limited in their ability to
detect millisecond and rapidly rotating pulsars by the low data
sampling rates and frequency resolution.  A current survey of the 
Magellanic Clouds is making use of a new data acquisition system,
BPSR, developed for the Parkes High Time Resolution Universe Legacy
Survey\footnote{http://astronomy.swin.edu.au/pulsar/?topic=hlsurvey}
which is currently undertaking a large-scale search for pulsars along
the Galactic plane. The parameters of this new system high-resolution
pulsar data acquisition system for our search of the Magellanic Clouds
are summarized in Table~\ref{tb:param}.  As can be seen, the time
resolution is improved by factors of 4 and 16 over the Crawford et
al.~and Manchester et al.~surveys, while the frequency resolution is
improved by a factor of 30.

With the MSP upper limits as references, we now create 
a sample distribution of the LMC and SMC MSP
populations, and attempt to detect the pulsars using 
the new survey parameters. Using 8400~s integrations, these simulations
predict that we will find at most 3 MSPs in each of the two 
Magellanic Clouds.  With a full survey of the LMC 
and SMC, the absence of MSP detections would lower the upper limits on 
the number of MSPs to 5,500 and 8,000, respectively.

A pilot survey of the SMC began in May 2009 using this new system.  The 
goal of this new survey is to detect or place stringent limits on the 
population of MSPs and giant-pulse emitting neutron stars 
in the Magellanic Clouds.  Following the survey pointings from \citet{ckm+01}, 
giant pulses from PSR B0540$-$69 have been observed, and three previously known 
pulsars have been redetected.  Processing of all data from this survey 
should be completed by August of 2010. 

We also briefly look at future predictions using the Square Kilometre Array 
(SKA).  Following the ``A'' implementation \citep{sks+09} of the SKA, we 
choose survey parameters that allow us to search the LMC and SMC for 
both normal and millisecond pulsars.  With a sky coverage per pointing of nearly 
250 deg$^2$, the LMC can be covered in only 4 pointings, while the 
SMC can be completely searched with only 3.  Using 10 minute integration 
times would keep the total survey time for either region under 1~hour, 
and results in 2,500 and 1,400 normal pulsar detections in the LMC 
and SMC respectively.  Additionally, this SKA survey could potentially 
detect 850 and 800 MSPs in the LMC and SMC, which is a considerable 
improvement over previous and current surveys of the Magellanic Clouds.

\section{CONCLUSIONS}\label{sec:conclusions}

We successfully reproduce the results from three past Magellanic Cloud 
surveys, giving us an accurate normal pulsar population for both the LMC 
and SMC.  These populations provide estimates of the number of normal pulsars 
located in the LMC and SMC of $(1.79 \pm 0.20) \times 10^4$ and $(1.09 \pm 0.16)
\times 10^4$ pulsars respectively.  Taking into account pulsars whose
beams do not intersect our line of sight, and the significant fraction
of pulsars which escape both the LMC and SMC, we find that the mean
radio pulsar birth rate is most likely to be in the range 0.5--1 pulsars
per century in either cloud. This is  consistent with historical
observations of supernovae in the Magellanic Clouds. We find that
a substantial population of radio pulsars will be ejected from the 
clouds and populate the local intergalactic medium. 

Extending our analysis to the MSP population yields upper limits 
of 15,000 for the LMC and 23,000 for the SMC.  If the current SMC survey 
at Parkes produces no MSP detections, that upper limit will be reduced 
to 8,000.  Consequently, we can realistically expect no more than 3 
MSPs to be detected in that survey.  Also, if the current SMC survey extends 
to the LMC, a maximum of 3 MSPs would be detected, but a lack of MSP 
detections would reduce the MSP upper limit to 5,500.

Our results here, along with the future results of the ongoing BPSR survey
will continue to constrain the number of pulsars, both normal and millisecond, 
found in the Magellanic Clouds.  With improved simulation techniques and 
more sensitive surveys, we will be able to model the true pulsar 
population with increased accuracy to allow for further research in this area. 
A logical extension of our work would be a full dynamical model of the
pulsar content in the Magellanic Clouds as has been recently carried
out for radio pulsars in our Galaxy \citep{fk06,rl10}.

\bibliographystyle{mn2e}
\bibliography{journals,modrefs,psrrefs,crossrefs}

\end{document}